# Ultra Wideband Impulse Radio Systems with Multiple Pulse Types[1]

Sinan Gezici[2], *Student Member, IEEE*, Zafer Sahinoglu[3], *Member, IEEE*,
Hisashi Kobayashi[2], *Life Fellow, IEEE*, and H. Vincent Poor[2], *Fellow, IEEE*

*Abstract*— In an ultra wideband (UWB) impulse radio (IR) system, a number of pulses, each transmitted in an interval called a "frame", is employed to represent one information symbol. Conventionally, a single type of UWB pulse is used in all frames of all users. In this paper, IR systems with multiple types of UWB pulses are considered, where different types of pulses can be used in different frames by different users. Both stored-reference (SR) and transmitted-reference (TR) systems are considered. First, the spectral properties of a multi-pulse IR system with polarity randomization is investigated. It is shown that the average power spectral density is the average of the spectral contents of different pulse shapes. Then, approximate closed-form expressions for the bit error probability of a multi-pulse SR-IR system are derived for RAKE receivers in asynchronous multiuser environments. The effects of both inter-frame interference (IFI) and multiple-access interference (MAI) are analyzed. The theoretical and simulation results indicate that SR-IR systems that are more robust against IFI and MAI than a "conventional" SR-IR system can be designed with multiple types of ultra-wideband pulses. Finally, extensions to multi-pulse TR-IR systems are briefly described.

*Index Terms*— Ultra-wideband (UWB), multi-pulse impulse radio (IR), stored-reference (SR), transmitted-reference (TR), performance analysis.

## I. Introduction

Ultra-wideband (UWB) technology holds great promise for a variety of applications such as short-range high-speed data transmission and precise location estimation. Commonly, impulse radio (IR) systems, which transmit very short pulses with a low duty cycle, are employed to implement UWB systems ([1]-[4]). In an IR system, a number $N_f$ of pulses are transmitted per symbol, and information is usually carried by the polarity of the pulses in a coherent system, or by the difference in the polarity of the pulses in a differentially-modulated system. In the former case, it is assumed that received pulse structure is known at the receiver and channel estimation can be performed; hence, RAKE receivers can be used to collect energy from different multipath components. Since the incoming signal structure is correlated by a locally stored reference (template) signal in this case, such a system is called a *stored-reference* (SR) system [5]. In the latter case, out of the $N_f$ pulses transmitted per information symbol, half of them are used as reference pulses, whereas the remaining half are used as data pulses. The relative polarity of the reference and the data pulses carries the information. Since the reference pulses to be used in the demodulation are transmitted to the receiver, such a system is called a *transmitted-reference* (TR) system [4]. In a TR system, there is no need for channel estimation since the reference and the data pulses are effected by the same channel, assuming that the channel is constant for a sufficiently long time interval, which is usually the case for UWB systems. On the other hand, a lower throughput is expected since half of the energy is used for non-information carrying pulses. Also since the transmitted reference is used as a noisy template at the receiver, more effective noise terms are generated.

Considering a conventional SR-IR system, a single type of UWB pulse is transmitted in all frames of all users [1]. In asynchronous multiuser environments, the autocorrelation function of the pulse becomes an important factor in determining the effects of inter-frame interference (IFI) and multiple-access interference (MAI) [6]. In order to reduce those effects, UWB pulses with fast decaying autocorrelation functions are desirable. However, such an autocorrelation function also results in a considerable decrease in the desired signal part of the receiver output in the presence of timing jitter [7]. Moreover, when there is an exact overlap between a pulse and an interfering pulse, the interference is usually very significant. Hence, there is not much flexibility in choosing the pulse shape in order to combat against interference effects. However, in SR-IR systems with multiple types of UWB pulses, the effects of interference can be mitigated by using different types of UWB pulses with good cross-correlation properties. Multi-pulse SR-IR systems have recently been proposed in [8]. However, there has been no theoretical analysis of such systems, in terms of their spectral properties and bit error probability (BEP) performance, and no quantitative investigation of the gains that can be obtained by multiple types of UWB pulses. In this paper, we consider this problem in an asynchronous multiuser environment and analyze the BEP performance of a generic RAKE receiver over frequency-selective channels. The results are valid for arbitrary numbers of UWB pulse types, and hence cover the single-pulse system as a special case. Moreover, we also briefly describe possible extensions of multi-pulse approach to TR-IR systems. However, no detailed analysis is given due to space limitations.

In addition to the performance analysis of the multi-pulse IR systems, the average power spectral density (PSD) of a generic multi-pulse IR signal is derived and a simple relationship between the Fourier transforms of the UWB pulses and the average PSD of the transmitted signal is obtained.

The remainder of the paper is organized as follows. Section II describes a generic transmitted signal model, which reduces to SR- and TR-IR systems as special cases. Then, Section III analyzes the spectral properties of this generic IR signal structure. After describing the received signal in Section IV, the performance of multi-pulse SR-IR systems employing RAKE receivers is analyzed in Section V and simulation results are given in Section VI. The concluding remarks are made and possible extensions are discusses in the last section.

[1]Manuscript received March 3, 2005; revised October 15, 2005. This research is supported in part by the National Science Foundation under grant ANI-03-38807, and in part by the New Jersey Center for Wireless Telecommunications. Part of this material was presented at the 39th Annual Conference on Information Science and Systems (CISS 2005).

[2]Department of Electrical Engineering, Princeton University, Princeton, NJ 08544, USA, Tel: (609) 258-6868, Fax: (609)258-2158, email: {sgezici,hisashi,poor}@princeton.edu

[3]Mitsubishi Electric Research Labs, 201 Broadway, Cambridge, MA 02139, USA, e-mail: zafer@merl.com

## II. TRANSMITTED SIGNAL MODEL

The transmitted signal from the $k$th user in a multi-pulse UWB-IR system can be expressed as

$$s^{(k)}(t) = \frac{1}{\sqrt{N_f}} \sum_{i=-\infty}^{\infty} \sum_{n=0}^{N_p-1} s_{i,n}^{(k)}(t), \quad (1)$$

where $N_f$ is the number of pulses transmitted per information symbol, $N_p$ is the number of different pulse types, and $s_{i,n}^{(k)}(t)$ represents the UWB pulses of type $n$ transmitted for the $i$th information symbol of user $k$. Note that the signal model in (1) can also represent cases in which the number of pulse types is less than $N_p$, by using the same pulses for different pulse indices. Also different users can have different ordering of the pulses in one period, which can be useful for reducing the effects of MAI. The number of pulses per symbol, $N_f$, is assumed to be an even multiple of $N_p$ for simplicity of notation and $s_{i,n}^{(k)}(t)$ is expressed as follows:

$$s_{i,n}^{(k)}(t) = \sum_{j=iN_f/(2N_p)}^{(i+1)\frac{N_f}{2N_p}-1} \Big\{ b_{1,j}^{(k)} d_{2jN_p+n}^{(k)} p_n^{(k)}\big(t - (2jN_p+n)T_f$$
$$- c_{2jN_p+n}^{(k)} T_c\big) + b_{2,j}^{(k)} d_{(2j+1)N_p+n}^{(k)} p_n^{(k)}\big(t - (2jN_p+n)T_f$$
$$- T_n^{(k)} - c_{(2j+1)N_p+n}^{(k)} T_c\big) \Big\}. \quad (2)$$

In (2), $p_n^{(k)}(t)$ is the UWB pulse of type $n$ for user $k$, $T_f$ is the frame interval, $T_c$ is the chip interval, and $T_n^{(k)}$ is the distance between the two pulses in a pair of type $n$ for the $k$th user, considering the pulses from a given type being grouped into pairs as shown in Figure 1. The time-hopping (TH) code for user $k$ is denoted by $c_j^{(k)}$, which is an integer taking values in the set $\{0, 1, \ldots, N_c-1\}$, with $N_c$ being the number of chips per frame, which prevents catastrophic collisions between different users. The polarity, or the spreading, code, $d_j^{(k)} \in \{-1, +1\}$, changes the polarity of the pulses, which smoothes the PSD of the transmitted signal [9] and provides robustness against MAI [10]. The information is represented by $b_{1,j}^{(k)}$ and $b_{2,j}^{(k)}$, which carry the same information for an SR system, and carry the information in the difference between their values for a TR system.

The general signal model in (2) can represent SR and TR systems as special cases:

### A. Stored Reference Impulse Radio

For the SR system, $b_{1,j}^{(k)} = b_{2,j}^{(k)} = b_{\lfloor 2N_p j/N_f \rfloor}^{(k)}$, $T_n^{(k)} = N_p T_f$ $\forall n, k$, and each frame has independent TH and polarity codes.

### B. Transmitted Reference Impulse Radio

For the TR system, $b_{1,j}^{(k)} = 1$ and $b_{2,j}^{(k)} = b_{\lfloor 2N_p j/N_f \rfloor}^{(k)}$. In other words, the first pulse in (2) is the reference pulse and the second one is the data pulse. As shown in Figure 1, this results in a structure in which the first $N_p$ pulses are the reference pulses, the next $N_p$ pulses are the data pulses, and which follows this alternating structure.

Note that $T_n^{(k)}$ can be chosen to be larger than $N_p T_f$ for the TR system if the TH sequence is constrained to a set

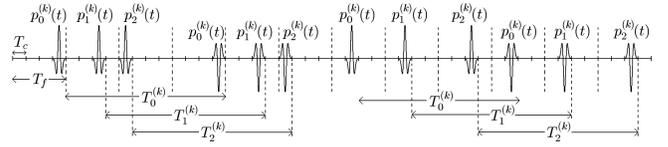

Fig. 1

TRANSMITTED SIGNAL FROM A MULTI-PULSE TR-IR SYSTEM, WHERE $N_f = 12$, $N_c = 4$, $T_n^{(k)} = \Delta T_c$ FOR $n = 0, 1, 2$ WITH $\Delta = 12$, AND THE TH SEQUENCE IS $\{3, 2, 0, 3, 2, 0, 1, 1, 2, 1, 1, 2\}$. FOR SIMPLICITY, NO POLARITY CODES ARE SHOWN (THAT IS, $d_j^{(k)} = 1 \,\forall j$), AND $b_{1,j}^{(k)} = 1$ AND $b_{2,j}^{(k)} = -1$.

$\{0, 1, \ldots, N_h-1\}$ with $N_h < N_c$. In this case, different values from the set $[N_p T_f, (N_p+1)T_f - N_h T_c]$ can be chosen for different users and/or different pulse types in order to provide extra robustness against the effects of interference.

Also each reference-data pulse pair has the same TH and polarity codes in order to facilitate simple delay and multiplication operation at the receiver. In other words, $d_{2jN_p+n}^{(k)} = d_{(2j+1)N_p+n}^{(k)}$, and $c_{2jN_p+n}^{(k)} = c_{(2j+1)N_p+n}^{(k)}$, for $j = iN_f/(2N_p), \ldots, (i+1)N_f/(2N_p) - 1$, $\forall\, i$, $n = 0, 1, \ldots, N_p - 1$.

## III. PSD OF MULTI-PULSE UWB-IR SYSTEMS

In order to evaluate the spectral properties of the transmitted signal, the (average) PSD of the signal must be calculated. Therefore, we first calculate the autocorrelation function of $s(t)$ as follows[4]:

$$\phi_{ss}(t+\tau, t) = \mathrm{E}\{s(t+\tau)s(t)\}$$
$$= \frac{1}{N_f} \sum_{i=-\infty}^{\infty} \sum_{n=0}^{N_p-1} \mathrm{E}\{s_{i,n}(t+\tau)s_{i,n}(t)\}, \quad (3)$$

where we employ the fact that the polarity codes are i.i.d. for different bit and pulse indices.

From (2), $\mathrm{E}\{s_{i,n}(t+\tau)s_{i,n}(t)\}$ can be calculated, after some manipulation, as

$$\mathrm{E}\{s_{i,n}(t+\tau)s_{i,n}(t)\} = \sum_{j=iN_f/N_p}^{(i+1)\frac{N_f}{N_p}-1} \mathrm{E}\Big\{ p_n\big(t+\tau - (jN_p+n)T_f$$
$$- c_{jN_p+n} T_c \big) p_n\big(t - (jN_p+n)T_f - c_{jN_p+n}T_c\big) \Big\}. \quad (4)$$

From (3) and (4), it is observed that $s(t)$ is not wide-sense stationary (WSS) since the autocorrelation function is not independent of $t$. However, note that $s(t)$ is a zero mean cyclostationary process since $\phi_{ss}(t+\tau, t)$ is periodic with a period of $N_p T_f$ [11]. Therefore, we can obtain the time-average autocorrelation function as

$$\bar{\phi}_{ss}(\tau) = \frac{1}{N_p T_f} \int_0^{N_p T_f} \phi_{ss}(t+\tau, t) dt$$
$$= \frac{1}{N_p T_f N_f} \sum_{n=0}^{N_p-1} \int_{-\infty}^{\infty} p_n(t+\tau)p_n(t)dt, \quad (5)$$

---
[4]We drop the user index $k$ in this section, for notational convenience.

the Fourier transform of which gives the average PSD as follows:

$$\Phi_{ss}(f) = \frac{1}{N_p T_s} \sum_{n=0}^{N_p-1} |W_n(f)|^2, \quad (6)$$

where $W_n(f)$ is the Fourier transform of $p_n(t)$.

Note from (6) that the average PSD of the signal is the average value of the squares of the Fourier transforms of the pulses. The dependence on the pulse spectra only is a consequence of the pulse-based polarity randomization, as considered for impulse radio systems in [9] and [12]. Moreover, we note that the multi-pulse system can have more flexibility in shaping the PSD by proper choice of the UWB pulses.

## IV. CHANNEL MODEL AND RECEIVED SIGNAL

We consider the following channel model for user $k$:

$$h^{(k)}(t) = \sum_{l=0}^{L-1} \alpha_l^{(k)} \delta(t - \tau_l^{(k)}), \quad (7)$$

where $\alpha_l^{(k)}$ and $\tau_l^{(k)}$ are, respectively, the fading coefficient and the delay of the $l$th path for user $k$.

Using the channel model in (7) and the transmitted signal in (1), the received signal can be expressed as

$$r(t) = \sum_{k=1}^{K} \sqrt{\frac{E_k}{N_f}} \sum_{i=-\infty}^{\infty} \sum_{n=0}^{N_p-1} \sum_{j=\frac{iN_f}{2N_p}}^{(i+1)\frac{N_f}{2N_p}-1} \left\{ b_{1,j}^{(k)} d_{2jN_p+n}^{(k)} u_n^{(k)}(t - (2jN_p+n)T_f - c_{2jN_p+n}^{(k)} T_c - \tau_0^{(k)}) + b_{2,j}^{(k)} d_{(2j+1)N_p+n}^{(k)} u_n^{(k)}(t - (2jN_p+n)T_f - T_n^{(k)} - c_{(2j+1)N_p+n}^{(k)} T_c - \tau_0^{(k)}) \right\} + \sigma n(t), \quad (8)$$

with

$$u_n^{(k)}(t) = \sum_{l=0}^{L-1} \alpha_l^{(k)} w_n^{(k)}(t - \tau_l^{(k)} + \tau_0^{(k)}), \quad (9)$$

where $w_n^{(k)}(t)$ is the received UWB pulse of type $n$ for user $k$, $E_k$ determines the received energy from user $k$, and $n(t)$ is a zero mean white Gaussian process with unit spectral density.

## V. ANALYSIS OF RAKE RECEIVERS FOR MULTI-PULSE SR-IR

Since $b_{1,j}^{(k)} = b_{2,j}^{(k)} = b_{\lfloor 2N_p j/N_f \rfloor}^{(k)}$, and each frame has independent TH and polarity codes for the SR-IR system, the received signal in (8) can be expressed, after some manipulation, as

$$r(t) = \sum_{k=1}^{K} \sqrt{\frac{E_k}{N_f}} \sum_{j=-\infty}^{\infty} b_{\lfloor j/N_f \rfloor}^{(k)} d_j^{(k)} u_j^{(k)}\left(t - jT_f - c_j^{(k)} T_c - \tau_0^{(k)}\right) + \sigma n(t), \quad (10)$$

with $u_j^{(k)}(t)$ given by (9). For the indices of the pulse types, such as in (9), the $modulo$ $N_p$ operation is implicitly assumed. In other words, for any $n \in \{0, 1, \ldots, N_p - 1\}$, $w_n(t) = w_{n+kN_p}(t)$ for all integers $k$.

We consider a generic RAKE receiver that can represent different combining schemes, such as equal gain or maximal ratio combining. It can be expressed as the correlation of the received signal in (8) with the following template signal, where we consider the 0th bit of user 1 without loss of generality:

$$s_{\text{temp}}^{(1)}(t) = \sum_{j=0}^{N_f-1} d_j^{(1)} v_j^{(1)}(t - jT_f - c_j^{(1)} T_c), \quad (11)$$

with

$$v_j^{(1)}(t) = \sum_{l=0}^{L-1} \beta_l w_j^{(1)}(t - \tau_l^{(1)}), \quad (12)$$

where $\beta_l$ denotes the RAKE combining coefficient for the $l$th path. We assume that $\tau_0^{(1)} = 0$, and $\tau_0^{(k)} \in [0, N_f T_f)$, for $k = 2, \ldots, K$, again without loss of generality. Note that for a partial or selective RAKE receiver [13], the combining coefficients for those paths that are not used are set to zero.

We assume that the delay spreads of the channels are not larger than one frame interval; that is $\tau_{L-1}^{(k)} \leq T_f$, $\forall k$. In other words, the frame interval is chosen to be sufficiently large so that the pulses in one frame can interfere only with those in the adjacent frames.

Using (10) and (11), the decision variable for detecting the 0th bit of user 1 can be obtained as:

$$Y = b_0^{(1)} \sqrt{\frac{E_1}{N_f}} \sum_{j=0}^{N_f-1} \phi_{u_j^{(1)} v_j^{(1)}}(0) + I + M + N, \quad (13)$$

with

$$\phi_{u_i^{(k)} v_j^{(l)}}(x) = \int u_i^{(k)}(t-x) v_j^{(l)}(t) dt, \quad (14)$$

where the first term in (13) is the desired signal part of the output, $I$ is the IFI, $M$ is the MAI, and $N$ is the output noise. For simplicity of notation, bit indices are not shown.

### A. Inter-frame Interference

The IFI occurs when a pulse of the desired user, user 1, in a given frame spills over to an adjacent frame due to multipath and consequently interferes with the pulse in that frame. The IFI for the 0th symbol can be expressed, using (10), (11) and (14), as the sum of the IFI to each frame of the template signal:

$$I = \sqrt{\frac{E_1}{N_f}} \sum_{j=0}^{N_f-1} \hat{I}_j, \quad (15)$$

where

$$\hat{I}_j = d_j^{(1)} \sum_{m \in \{\pm 1\}} d_{j+m}^{(1)} b_{\left\lfloor \frac{j+m}{N_f} \right\rfloor}^{(1)} \phi_{u_{j+m}^{(1)} v_j^{(1)}}\left(mT_f + (c_{j+m}^{(1)} - c_j^{(1)})T_c\right). \quad (16)$$

Note that due to the assumption on the delay spreads of the channels, the IFI occurs only between adjacent frames.

The asymptotic distribution of the IFI in (15) is given by the following proposition:

*Proposition 5.1:* Consider a random TH SR-IR system with pulse-based polarity randomization, which employs $N_p > 1$ different UWB pulses. Then, the IFI at the output of the RAKE receiver, expressed by (13), is asymptotically distributed as

$$I \sim \mathcal{N}\left(0, \frac{E_1}{N_p N_c^2} \sum_{n=0}^{N_p-1} \left[\sigma_{\text{IFI},1}^2(n) + 2\sigma_{\text{IFI},2}^2(n)\right]\right), \quad (17)$$

as $\frac{N_f}{N_p} \longrightarrow \infty$, where

$$\sigma_{\text{IFI},1}^2(n) = \sum_{l=1}^{N_c} l \left( \phi_{u_{n+1}^{(1)} v_n^{(1)}}^2 (lT_c) + \phi_{u_{n-1}^{(1)} v_n^{(1)}}^2 (-lT_c) \right),$$

$$\sigma_{\text{IFI},2}^2(n) = \sum_{l=1}^{N_c} l \, \phi_{u_{n+1}^{(1)} v_n^{(1)}} (lT_c) \phi_{u_n^{(1)} v_{n+1}^{(1)}} (-lT_c). \quad (18)$$

*Proof:* See Appendix A.

Due to the FCCs regulation on peak to average ratio (PAR), $N_f$ cannot be chosen very small in practice. Since we transmit a certain amount of energy in a constant symbol interval, as $N_f$ gets smaller, the signal becomes peakier [7]. Therefore, the approximation for large $N_f/N_p$ can be quite accurate for real systems depending on the number of pulse types and the other system parameters.

From Proposition 5.1, the following result for a double-pulse system can be obtained.

*Corollary 5.1:* Consider a random TH SR-IR system with pulse-based polarity randomization, where the UWB pulses $w_0(t)$ and $w_1(t)$, which are both even functions, are transmitted alternately. For this system, the IFI in (13) is approximately distributed as follows for large $N_f$:

$$I \sim \mathcal{N} \left( 0, \frac{E_1}{N_c^2} \sum_{l=1}^{N_c} l \left[ \phi_{u_0^{(1)} v_1^{(1)}} (-lT_c) + \phi_{u_0^{(1)} v_1^{(1)}} (lT_c) \right]^2 \right). \quad (19)$$

The distribution of the IFI for the case where a single UWB pulse $w_0(t)$ is used in all the frames is given by [14]

$$I \sim \mathcal{N} \left( 0, \frac{E_1}{N_c^2} \sum_{l=1}^{N_c} l \left[ \phi_{u_0^{(1)} v_0^{(1)}} (-lT_c) + \phi_{u_0^{(1)} v_0^{(1)}} (lT_c) \right]^2 \right). \quad (20)$$

Note that in an IFI-limited scenario, the autocorrelation function of the UWB pulse is the determining factor for a single-pulse system. However, for the system using multiple types of UWB pulses, the IFI is determined by the cross-correlations of different pulses. Note that it is possible to design the pulses so that they are orthogonal and their cross-correlations decay quickly, e.g. modified Hermite pulses (MHPs) [15]. However, the autocorrelation function always causes large values when there is an exact overlap of the multipath components. Also a rapidly decaying autocorrelation function, which is good for combatting the IFI, may not be very desirable since small timing jitter in the system could result in a significant loss in the desired signal part of the decision variable. Therefore, the multi-pulse IR system is expected to have better IFI rejection capability than the single-pulse system. For example, for a system with $N_f = 20$, $N_c = 30$ and $L = 20$, the power of the IFI is reduced by about 30% by using the 4th and 5th order MHPs instead of using the 4th order MHP only.

### B. Multiple-Access Interference

Consider the MAI term $M$ in (13), which is the sum of the interference terms from $(K-1)$ users; that is, $M = \sum_{k=2}^{K} M^{(k)}$, where $M^{(k)}$ can be expressed as $M^{(k)} = \sqrt{\frac{E_k}{N_f}} \sum_{j=0}^{N_f - 1} \hat{M}_j^{(k)}$, with $\hat{M}_j^{(k)}$ denoting the MAI from user $k$ to the $j$th frame of the first user. From (10), (11) and (14), $\hat{M}_j^{(k)}$ can be expressed as

$$\hat{M}_j^{(k)} = d_j^{(1)} \sum_{m=-\infty}^{\infty} d_m^{(k)} b_{\lfloor m/N_f \rfloor}^{(k)} \phi_{u_m^{(k)} v_j^{(1)}} \Big( (m-j)T_f + (c_m^{(k)} - c_j^{(1)})T_c + \tau_0^{(k)} \Big), \quad (21)$$

where $\tau_0^{(k)}$ denotes the amount of asynchronism between user $k$ and the user of interest, user 1, since we assume $\tau_0^{(1)} = 0$.

For a given value of $\tau_0^{(k)}$, the distribution of the MAI from user $k$ can be obtained approximately from the following proposition:

*Proposition 5.2:* Consider a random TH SR-IR system with pulse-based polarity randomization, which employs $N_p$ different types of UWB pulses. Then, the MAI from user $k$, $M^{(k)}$, given $\tau_0^{(k)}$, is asymptotically distributed as follows

$$M^{(k)} | \tau_0^{(k)} \sim \mathcal{N} \left( 0, \frac{E_k}{N_p N_c^2} \sum_{n=0}^{N_p - 1} \sigma_{\text{MAI},k}^2 (n, \tau_0^{(k)}) \right), \quad (22)$$

as $\frac{N_f}{N_p} \longrightarrow \infty$, where

$$\sigma_{\text{MAI},k}^2(n, \tau_0^{(k)}) = \sum_{m \in \mathcal{A}} \sum_{l=-(N_c - 1)}^{N_c - 1} (N_c - |l|)$$
$$\times \phi_{u_m^{(k)} v_n^{(1)}}^2 \Big( ((m-n)N_c + l) T_c + \tau_0^{(k)} \Big), \quad (23)$$

with $\mathcal{A} = \left\{ \left\lceil n - 2 + \frac{1}{N_c} - \frac{\tau_0^{(k)}}{T_f} \right\rceil, \ldots, \left\lfloor n + 2 - \frac{1}{N_c} - \frac{\tau_0^{(k)}}{T_f} \right\rfloor \right\}$.

*Proof:* The proof is similar to that of Proposition 5.1, and is omitted due to space limitations.

Note that the Gaussian approximation in Proposition 5.2 is different from the standard Gaussian approximation (SGA) used in analyzing a system with many users ([17]-[19]). Proposition 5.2 states that when the number of *pulses* per information symbol is large compared to the number of different pulse types, the MAI from an interfering user is approximately distributed as a Gaussian random variable. This idea is similar to the improved Gaussian approximation approach in [16], where the large processing gain of a CDMA system leads to normally distributed MAI conditioned on some systems parameters.

Denote the amount of asynchronism between user $k$ and user 1 as $\tau_0^{(k)} = \lfloor \tau_0^{(k)} / T_c \rfloor T_c + \epsilon_k$, where $\epsilon_k \in [0, T_c)$. When a single type of UWB pulse is employed in the system, it can be shown from Proposition 5.2 that the distribution of $M^{(k)}$ is given by the following result:

*Corollary 5.2:* Consider a random TH SR-IR system with pulse-based polarity randomization, where the UWB pulse $w_0(t)$ is employed in all frames of all users. Then, the conditional distribution of the MAI from user $k$ is given by

$$M^{(k)} | \tau_0^{(k)} \sim \mathcal{N} \left( 0, \frac{E_k}{N_c} \sum_{l=-N_c}^{N_c - 1} \phi_{u_0^{(k)} v_0^{(1)}}^2 (lT_c + \epsilon_k) \right). \quad (24)$$

Note from Corollary 5.2 that the distribution of $M^{(k)}$ depends on $\epsilon_k$, instead of $\tau_0^{(k)}$, for a single-pulse system. This is because the probability that a given pulse of the desired user collides with the pulses of user $k$ is the same for all delays $\tau_0^{(k)}$ with identical $\epsilon_k$ values, due to the random TH codes, and the same amount of average interference occurs when the same pulses are used in all frames.

Denote $\boldsymbol{\tau} = [\tau_0^{(2)} \cdots \tau_0^{(K)}]$. Then, given $\boldsymbol{\tau}$, the distribution of the total MAI $M$ in (13) can be approximated by

$$M|\boldsymbol{\tau} \sim \mathcal{N}\left(0, \frac{1}{N_p N_c^2} \sum_{k=2}^{K} \sum_{n=0}^{N_p-1} E_k \sigma_{\text{MAI},k}^2(n, \tau_0^{(k)})\right), \quad (25)$$

for large $N_f/N_p$, where $\sigma_{\text{MAI},k}^2(n, \tau_0^{(k)})$ is as in (23). Note that it is not necessary to have a large number of users, or equal energy interferers (perfect power control), for the expression in (25) to be accurate. The only requirement is to have a large ratio between the number of pulses per symbol and the number of pulse types.

When the delays of the interferers are unknown and/or an average performance measure is to be obtained, then each interferer is assumed to have a uniformly distributed delay with respect to the desired user; that is, $\tau_0^{(k)} \sim [0, N_f T_f)$, $\forall k$. In this case, the performance measure, such as the BEP expression, needs to be averaged over the distribution of $\boldsymbol{\tau}$.

### C. Output Noise

The output noise $N$ in (13) is distributed as $\mathcal{N}\left(0, \sigma^2 \int |s_{\text{temp}}^{(1)}(t)|^2 dt\right)$. Using the expression in (11) for $s_{\text{temp}}^{(1)}(t)$, we can approximate the distribution of $N$ for an SR-IR system with a single UWB pulse $w_0(t)$ as $N \sim \mathcal{N}\left(0, N_f \sigma^2 \phi_{v_0^{(1)}}(0)\right)$, for large values of $N_f$, where $\phi_{v_j^{(k)}}(x) = \int v_j^{(k)}(t-x) v_j^{(k)}(t) dt$ is the autocorrelation function of $v_j^{(k)}(t)$.

Similarly, for an SR-IR system employing $N_p$ types of pulses, we obtain the approximate distribution of $N$ as $N \sim \mathcal{N}\left(0, \sigma^2 \frac{N_f}{N_p} \sum_{n=0}^{N_p-1} \phi_{v_n^{(1)}}(0)\right)$, for large $N_f/N_p$.

### D. Bit Error Probability

Using the results in the previous sections, we can obtain an approximate BEP expression for the multi-pulse SR-IR system as follows:

$$P_e(\boldsymbol{\tau}) \approx Q\left(\frac{\sqrt{\frac{E_1}{N_p}} \sum_{n=0}^{N_p-1} \phi_{u_n^{(1)} v_n^{(1)}}(0)}{\sqrt{\sum_{n=0}^{N_p-1} \left[\sigma_{\text{IFI}}^2(n) + \sigma_{\text{MAI}}^2(n, \boldsymbol{\tau}) + \sigma^2 \phi_{v_n^{(1)}}(0)\right]}}\right), \quad (26)$$

for large $N_f/N_p$, where

$$\sigma_{\text{IFI}}^2(n) = \frac{E_1}{N_c N}[\sigma_{\text{IFI},1}^2(n) + 2\sigma_{\text{IFI},2}^2(n)], \quad (27)$$

$$\sigma_{\text{MAI}}^2(n, \boldsymbol{\tau}) = \frac{1}{N_c N} \sum_{k=2}^{K} E_k \sigma_{\text{MAI},k}^2(n, \tau_0^{(k)}), \quad (28)$$

$\boldsymbol{\tau} = [\tau_0^{(2)} \cdots \tau_0^{(K)}]$, $N = N_c N_f$ is the total processing gain of the system, $\sigma_{\text{IFI},1}^2(n)$ and $\sigma_{\text{IFI},1}^2(n)$ are as in (18) and $\sigma_{\text{MAI},k}^2(n, \tau_0^{(k)})$ is as in (23).

If we consider a synchronous scenario, where $\tau_0^{(k)} = 0$, for $k = 1, 2, \ldots, K$, then the unconditional BEP is given by $P_e^{\text{sync}} = P_e(\mathbf{0})$, with $P_e(\boldsymbol{\tau})$ being given by (26).

For an asynchronous system, we assume that $\tau_0^{(2)}, \ldots, \tau_0^{(K)}$ are i.i.d. distributed as $\mathcal{U}[0, T_s)$, where $T_s = N_f T_f$ is the symbol interval. Hence, the unconditional BEP can be obtained by

$$P_e^{\text{async}} = \frac{1}{T_s^{K-1}} \int_0^{T_s} \cdots \int_0^{T_s} P_e(\boldsymbol{\tau}) d\tau_0^{(2)} \ldots d\tau_0^{(K)}. \quad (29)$$

Due to the periodicity of the pulse structure, we can show that it is enough to average over an interval of length $N_p T_f$ instead of $N_f T_f$. Hence, $P_e^{\text{async}}$ can be expressed as

$$P_e^{\text{async}} = \frac{1}{(N_p T_f)^{K-1}} \int_0^{N_p T_f} \cdots \int_0^{N_p T_f} P_e(\boldsymbol{\tau}) d\tau_0^{(2)} \ldots d\tau_0^{(K)}. \quad (30)$$

In order to calculate $P_e^{\text{async}}$, numerical techniques or Monte-Carlo simulations can be used. For example, by generating $N_m$ vectors according to the uniform distribution in $[0, N_p T_f)^{K-1}$, we can approximate $P_e^{\text{async}}$ by Monte-Carlo simulations as $P_e^{\text{async}} = \frac{1}{N_m} \sum_{i=1}^{N_m} P_e(\boldsymbol{\tau}_i)$, where $\boldsymbol{\tau}_i$ denotes the $i$th random vector of interferer delays.

Note that the BEP expression in (30) becomes more accurate as $N_f/N_p$ gets larger, without the need for large number of users or equal energy interferers, which are needed for accurate BEP using the SGA. The SGA directly calculates the average value of the variance of the total MAI instead of averaging over a conditional BEP expression in (26). In other words, $P_e^{\text{async}}$ is approximated by the expression in (26) with the only change of using $\frac{1}{N_p T_f} \int_0^{N_p T_f} \sigma_{\text{MAI},k}^2(n, \tau_0^{(k)}) d\tau_0^{(k)}$ instead of $\sigma_{\text{MAI},k}^2(n, \tau_0^{(k)})$. Of course, this expression is easier to evaluate than the expression in (30), especially when there is a large number of users. Therefore, in such a case, the SGA might be preferred if the users' power levels are not very different. But for systems with small numbers of interferers, such as an IEEE 802.15.3a personal area network (PAN), the expression in (30) is not very difficult to evaluate and can result in more accurate BEP evaluations.

Now consider the case in which a single type of UWB pulse $w_0(t)$ is employed for all users. The BEP expression for this scenario can be obtained from (13), (20), (24), and Section V-C as

$$P_e(\boldsymbol{\epsilon}) \approx Q\left(\frac{\sqrt{E_1} \phi_{u_0^{(1)} v_0^{(1)}}(0)}{\sqrt{\sigma_{\text{IFI}}^2 + \sigma_{\text{MAI}}^2(\boldsymbol{\epsilon}) + \sigma^2 \phi_{v_0^{(1)}}(0)}}\right), \quad (31)$$

for large $N_f/N_p$, where

$$\sigma_{\text{IFI}}^2 = \frac{E_1}{N_c N} \sum_{l=1}^{N_c} l \left[\phi_{u_0^{(1)} v_0^{(1)}}(-l T_c) + \phi_{u_0^{(1)} v_0^{(1)}}(l T_c)\right]^2,$$

$$\sigma_{\text{MAI}}^2(\boldsymbol{\epsilon}) = \frac{1}{N} \sum_{k=2}^{K} \sum_{l=-N_c}^{N_c-1} E_k \phi_{u_0^{(k)} v_0^{(1)}}^2(l T_c + \epsilon_k), \quad (32)$$

and $\boldsymbol{\epsilon} = [\epsilon_2 \cdots \epsilon_K]$ characterizes the asynchronism between the interfering users and the desired user in $modulo\ T_c$ arithmetic. Similar to the multi-pulse case, the unconditional BEP is given by $P_e^{\text{sync}} = P_e(\mathbf{0})$, with $P_e(\boldsymbol{\epsilon})$ being as in (31), for the synchronous case, and by

$$P_e^{\text{async}} \approx \frac{1}{T_c^{K-1}} \int_0^{T_c} \cdots \int_0^{T_c} P_e(\boldsymbol{\tau}) d\tau_0^{(2)} \ldots d\tau_0^{(K)}, \quad (33)$$

for the asynchronous case.

From the closed-form BEP expressions for multi-pulse and single-pulse systems, we can observe that the IFI and MAI terms depend on the autocorrelation function of a single pulse for single-pulse systems, whereas they depend also on

6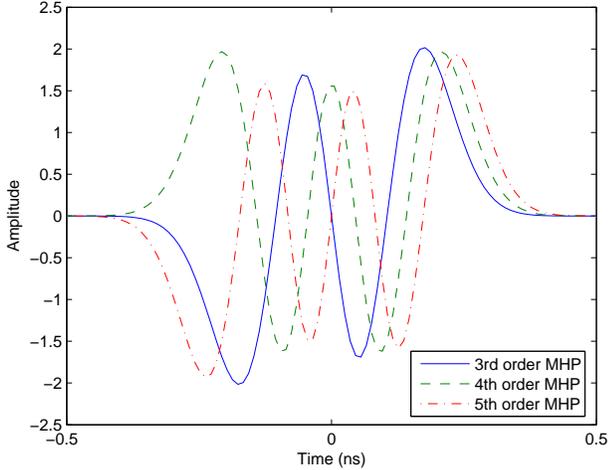

Fig. 2

THE 3RD, 4TH AND 5TH ORDER MODIFIED HERMITE PULSES (MHPs).

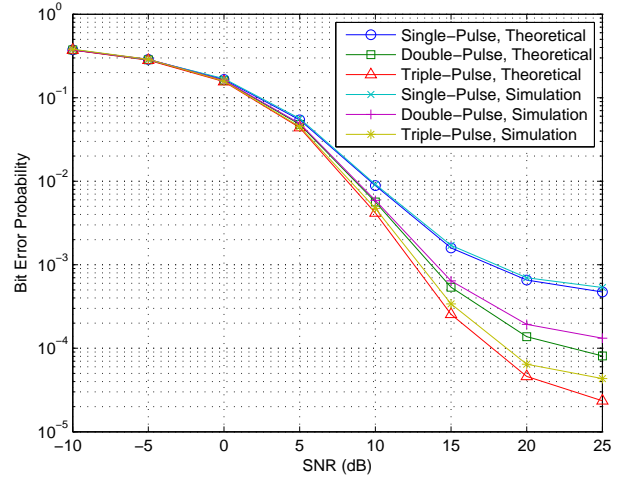

Fig. 3

THE BEP PERFORMANCE OF THE SINGLE-PULSE, DOUBLE-PULSE, AND TRIPLE-PULSE SR-IR SYSTEMS.

the cross-correlations of different pulse types for multi-pulse systems, which suggests that more flexibility in combatting the effects of the IFI and MAI is present in multi-pulse systems. In other words, by design of UWB pulses with good cross-correlation properties, it is possible to mitigate the IFI and MAI to a larger extent, as will be investigated in the next section.

## VI. SIMULATION RESULTS

In this section, we compare the BEP performance of single-pulse, double-pulse and a triple-pulse SR-IR systems. In the triple-pulse system, each user transmits the 3rd, 4th and 5th order MHPs [8] alternately, whereas the double-pulse system employs the 3rd and 4th order MHPs. For the single-pulse system, the 3rd order MHP is transmitted in all the frames (see Figure 2). The system parameters are $K = 5$ users, $N_f = 18$ frames per symbol, $N_c = 30$ chips per frame, and $T_c = 1$ ns. We consider an MAI-limited scenario, where the received energy of each interferer is 18.75dB more than that of the user of interest. All the channels have $L = 20$ taps, which are generated independently according to a channel model with exponentially decaying ($\mathrm{E}\{|\alpha_l|^2\} = \Omega_0 e^{-\lambda l}$) log-normal fading ($|\alpha_l| \sim \mathcal{LN}(\mu_l, \sigma^2)$) channel amplitudes, random signs for channel taps, and exponential distribution for the path arrivals with a mean $\hat{\mu}$. The channel parameters are $\lambda = 0.5$, $\sigma^2 = 1$, and $\hat{\mu} = 1.5$ ns, and $\mu_l$ can be calculated from $\mu_l = 0.5 \left[\ln(\frac{1-e^{-\lambda}}{1-e^{-\lambda L}}) - \lambda l - 2\sigma^2\right]$, for $l = 0, 1, \ldots, L-1$.

Figure 3 shows the BEP performance of all-RAKE receivers [13] for the single, double and triple-pulse systems. Both the theoretical and the simulation results are shown, which are in quite good agreement. From the figure, the effects of multiple pulse types on reducing the interference, hence the BEP, are observed. As the number of pulse types increases, more gain is obtained. Further gains can be obtained by using a larger number of UWB pulse types and/or MHPs that are several orders apart [8]. Also, the theoretical results are more accurate for smaller number of pulse types, $N_p$, since the asymptotic results in Section V assume large $N_f/N_p$ values.

## VII. CONCLUSIONS AND EXTENSIONS

In this paper, we have considered multi-pulse IR systems. First, we have introduced a generic model for an IR signal, which can represent an SR or a TR signal as special cases. Using this model, we have investigated the average PSD of the transmitted signal, which is important considering the power limitations imposed by the FCC. Then, we have provided a detailed BEP analysis for a multi-pulse SR-IR system, considering the effects of both the IFI and the MAI, and performed simulation studies to verify the theory.

The multi-pulse approach can be extended to TR-IR systems as well in order to mitigate the effects of interference. In this case, different pulse types can be transmitted next to each other as shown in Figure 1. If the same delay between the reference and data pulses is used for all pulse types, then a conventional TR receiver can be employed [4]. If different delays between reference and data pulses are employed for different pulse types, the receiver needs to perform $N_p$ parallel delay-and-multiply operations and combine the outputs of different branches. Theoretical and simulation studies are necessary to quantify the possible improvements by the use of multiple UWB pulses.

## APPENDIX

### A. Proof of Proposition 5.1

Let $N_f = N_r N_p$. Then, (15) can be expressed as $I = \sqrt{\frac{E_1}{N_r}} \sum_{j=0}^{N_r-1} \tilde{I}_j$, where $\tilde{I}_j = \frac{1}{\sqrt{N_p}} \sum_{n=0}^{N_p-1} \hat{I}_{jN_p+n}$ with $\hat{I}_j$ being given by (16). It can be shown, from (16), that $\mathrm{E}\{\tilde{I}_j\} = 0$, $\forall j$ due to the i.i.d. random polarity codes. Also from the $N_p \geq 2$ assumption in the proposition, it is straightforward to show that $\mathrm{E}\{\tilde{I}_j \tilde{I}_{j+l}\} = 0$ for $l \geq 2$, since $\tilde{I}_j$ and $\tilde{I}_{j+l}$ include terms with polarity codes of different indices, which are independent and zero mean by assumption. Hence, $\{\tilde{I}_j\}_{j=0}^{N_r-1}$ forms a zero mean 1-dependent sequence[5].

We employ the following central limit argument for dependent sequences to approximate the distribution of the IFI:

*Theorem 1:* [20] Consider a stationary d-dependent sequence of random variables $X_1, X_2, ...$ with $\mathrm{E}\{X_1\} = 0$ and $\mathrm{E}\{|X_1|^3\} < \infty$. If $S_n = X_1 + \ldots + X_n$, then $\frac{S_n}{\sqrt{n}} \xrightarrow{d} \mathcal{N}(0, \sigma^2)$, as $n \longrightarrow \infty$, where $\sigma^2 = \mathrm{E}\{X_1^2\} + 2\sum_{k=1}^{d} \mathrm{E}\{X_1 X_{1+k}\}$.

In order to apply the results of the theorem we first calculate the variance of $\tilde{I}_j$:

$$\mathrm{E}\{\tilde{I}_j^2\} = \frac{1}{N_p} \sum_{n_1=0}^{N_p-1} \sum_{n_2=0}^{N_p-1} \mathrm{E}\{\hat{I}_{jN_p+n_1} \hat{I}_{jN_p+n_2}\} \quad (34)$$

$$= \frac{1}{N_p} \sum_{n=0}^{N_p-1} \mathrm{E}\{\hat{I}_{jN_p+n}^2\} + \sum_{|n_1-n_2|=1} \mathrm{E}\{\hat{I}_{jN_p+n_1} \hat{I}_{jN_p+n_2}\},$$

where the second equality is obtained from (16) by using the fact that the polarity codes form an i.i.d. sequence. Then, after some manipulation, $\mathrm{E}\{\hat{I}_j^2\}$ can be expressed as

$$\mathrm{E}\{\hat{I}_j^2\} = \frac{1}{N_c^2} \sum_{l=1}^{N_c} l \left[ \phi^2_{u_{j-1}^{(1)} v_j^{(1)}}(-lT_c) + \phi^2_{u_{j+1}^{(1)} v_j^{(1)}}(lT_c) \right], \quad (35)$$

and $\mathrm{E}\{\hat{I}_j \hat{I}_{j+1}\}$ can be expressed as

$$\mathrm{E}\{\hat{I}_j \hat{I}_{j+1}\} = \frac{1}{N_c^2} \sum_{l=1}^{N_c} l \phi_{u_{j+1}^{(1)} v_j^{(1)}}(lT_c) \phi_{u_j^{(1)} v_{j+1}^{(1)}}(-lT_c). \quad (36)$$

In obtaining (35) and (36), we have used the expression in (16) and the facts that the polarity codes are randomly distributed in $\{-1, +1\}$ and the TH codes in $\{0, 1, \ldots, N_c - 1\}$.

Now considering the correlation between the adjacent terms of $\{\tilde{I}_j\}_{j=0}^{N_r-1}$, the following expression can be obtained:

$$\mathrm{E}\{\tilde{I}_j \tilde{I}_{j+1}\} = \frac{1}{N_p} \mathrm{E}\{\hat{I}_{(j+1)N_p-1} \hat{I}_{(j+1)N_p}\}. \quad (37)$$

Theorem 1 can be invoked for $\{\tilde{I}_j\}_{j=0}^{N_r-1}$, which results in $I \sim \mathcal{N}\left(0, E_1[\mathrm{E}\{\tilde{I}_j^2\} + 2\mathrm{E}\{\tilde{I}_j \tilde{I}_{j+1}\}]\right)$. Then, from (34)-(37), the distribution of $I$ can be approximated as in (17), as $N_r \longrightarrow \infty$.

**Sinan Gezici** received the B.S. degree from Bilkent University, Turkey in 2001, and the M.A. degree from Princeton University in 2003. He is currently working toward the Ph.D. degree at the Department of Electrical Engineering at Princeton University.

His research interests are in the communications and signal processing fields. Currently, he has a particular interest in synchronization, positioning, performance analysis and multiuser aspects of UWB communications.

**Zafer Sahinoglu** received his B.S. in EE from Gazi University, Turkey, M.S. in biomedical engineering and PhD (with awards) in EE from New Jersey Institute of Technology (NJIT) in 1998 and 2001 respectively. He worked at AT&T Shannon Research Labs in 1999, and has been with MERL since 2001. His current research interests include NWK and MAC layer issues in wireless sensor networks and impulse radio ultrawideband precision ranging and positioning. He has co-authored a book-chapter on UWB geo-location, and has been author and co-author of more than 30 conference and journal articles; and has provided significant contributions to emerging MPEG-21 standards on mobility modeling and characterization for multimedia service adaptation, to ZigBee on data broadcasting, routing and application profile development, to IEEE 802.15.4a standards on precision ranging. He is currently the chair of ZigBee Industrial Plant Monitoring Application Profile Task Group. He holds one European and six US patents, and has 20 pending.

**Hisashi Kobayashi** is the Sherman Fairchild University Professor of Electrical Engineering and Computer Science at Princeton University since 1986, when he joined the Princeton faculty as the Dean of the School of Engineering and Applied Science (1986-91). His current research fields include: network security protocols, ultra wideband (UWB) communications, wireless geolocation, hidden semi-Markov model (HSMM) and its computation algorithms. He is authoring with Brian L. Mark a graduate textbook "Modeling and Analysis: Foundations of System Performance Evaluation" (to be published by Prentice Hall, 2006).

---

[5] A sequence $\{X_n\}_{n \in \mathbb{Z}}$ is called a $D$-dependent sequence, if all finite dimensional marginals $(X_{n_1}, ..., X_{n_i})$ and $(X_{m_1}, ..., X_{m_j})$ are independent whenever $m_1 - n_i > D$.

He is a member of the Engineering Academy of Japan, a Life Fellow of IEEE, and a Fellow of IEICE of Japan. He received the 2005 Eduard Rhein Technology Award for his pioneering work on high density digital recording technique, widely known as PRML (Partial Response channel coding and Maximum Likelihood sequence detection). He was the recipient of the Humboldt Prize from Germany (1979), the IFIP's Silver Core Award (1981), and two IBM Outstanding Contribution Awards. He received his Ph.D. degree (1967) from Princeton University and BE and ME degrees (1961, 63) from the University of Tokyo, all in Electrical Engineering.

**H. Vincent Poor** (S'72, M'77, SM'82, F'77) received the Ph.D. degree in EECS from Princeton University in 1977. From 1977 until 1990, he was on the faculty of the University of Illinois at Urbana-Champaign. Since 1990 he has been on the faculty at Princeton, where he is the George Van Ness Lothrop Professor in Engineering. Dr. Poor's research interests are in the areas of statistical signal processing and its applications in wireless networks and related fields. Among his publications in these areas is the recent book Wireless Networks: Multiuser Detection in Cross-Layer Design (Springer: New York, NY, 2005).

Dr. Poor is a member of the National Academy of Engineering and is a Fellow of the American Academy of Arts and Sciences. He is also a Fellow of the Institute of Mathematical Statistics, the Optical Society of America, and other organizations. In 1990, he served as President of the IEEE Information Theory Society, and he is currently serving as the Editor-in-Chief of these Transactions. Recent recognition of his work includes the Joint Paper Award of the IEEE Communications and Information Theory Societies (2001), the NSF Director's Award for Distinguished Teaching Scholars (2002), a Guggenheim Fellowship (2002-03), and the IEEE Education Medal (2005).